**Towards an Ideometrics-Based General Theory of Human Progress**

Igor Rudan[1,2,3] and Steven Kerr[4]

[1]Centre for Global Health, Usher Institute, The University of Edinburgh, UK
[2]Nuffield Department of Primary Care Health Sciences, Oxford University, UK
[3]Green Templeton College, Oxford University, UK
[4]Centre for Medical Informatics, Usher Institute, The University of Edinburgh, UK

Address for Correspondence:

Professor Igor Rudan, FRSE, MAE, MEASA
Centre for Global Health, Usher Institute
5–7 Little France Road,
Edinburgh EH16 4UX,
Scotland, UK
E-mail: Igor.Rudan@ed.ac.uk; Igor.Rudan@phc.ox.ac.uk

**ABSTRACT**

This paper proposes ideometrics as the foundation for a generalised and potentially testable theory of human progress and civilisational progress, thus linking ideometrics to studies in economics and history. Building on prior work that conceptualises the human brain as a "sensor of ideas", the paper integrates diverse disciplinary perspectives, from economics, decision theory and evolution to philosophy of science and information theory, into a single, closed-loop framework. Within this framework, human progress is understood not primarily through outcomes such as wealth, health, or technological advancement, but through the dynamic process by which ideas are generated, evaluated, prioritised, and implemented to shape future states. The paper advances a formal definition of "human progress" as a measurable improvement in the ability of individuals and societies to generate, evaluate, prioritise, and implement ideas in a way that increasingly aligns prioritised ideas - under scarcity of human capacity, energy, time and resources - with those that truly lead to preferred future states, given available information and uncertainty. It further introduces the "Ideometric Index of Human Progress" (IIHP), a composite measure capturing the quality of idea generation (G), accuracy of their evaluation (E), efficiency of their prioritisation (P), and effectiveness of their implementation (Ie). It then shows that, although IIHP can serve to evaluate the quality of the process of ideometrics, the true progress in the future will only be realised if there is good alignment between the "perceived future value of ideas" and their true, realised future value, which can be assessed at different future points as "outcome monitoring" (O). This formulation shifts the analytical focus from static outcomes to the quality of evaluating ideas, thereby offering a novel lens for understanding both progress and regress. While acknowledging theoretical and practical challenges, the paper argues that approximate measurement through observable proxies can enable first empirical testing. The concept can also be extended to long periods of history through the "Ideometric Index of Civilisational Progress" (IICP), where additional parameters of successful documentation of outcomes (D) and successful intergenerational transmission of gathered knowledge (T) are added. IICP exposes fragility of civilisations, because any effective action to reduce quality of either G, E, P, Ie, O, D or T variables is sufficient to interfere with civilisational progress and eventually lead to regress. By transforming "ideas" into measurable units of analysis and emphasising the improvement of selection processes, rather than outcomes alone, ideometrics offers a potentially transformative approach to understanding human progress. This has practical implications for funding new research, supporting start-up companies, developing strategies for existing companies, formulating governmental policies, and advancing artificial intelligence systems.

**Introduction**

We previously explained an integrated multi-layer system, linking biology (human brain as a "sensor of ideas") [**1**], epistemology ("perceived value of acquired information" shapes evaluation of ideas) [**2**], methodology (the Child Health and Nutrition Research Initiative (CHNRI) method quantifies crowdsourced human collective opinion to "measure ideas" through its multi-criteria approach) [**3**], new framework for scientific inquiry ("ideometrics" unifies all idea-related processes, from generation, to evaluation, prioritisation, followed by implementation) [**4**], and evolution (where the entire global science, performed through "global research system" over time and space, and recently supported by AI, serves as a large "ideometric engine") [**5**].

In this paper, we aim to demonstrate that these five pieces do not just "fit together", but rather integrate to form a closed theoretical loop. The loop starts with ideas. From the human brain's sensory perspective, "ideas" are competing possibilities of purposeful activities that, if followed, would be expected to result in an alternative version of the future [**1**]. This component of invested energy, time, and expectations of the altered future based on an idea, is quite important to note: after the ideas are either generated, or perceived, they are evaluated by human brains.

The evaluation process may occur quickly, or over a longer period, as Kahneman suggested [**6**]. It is based on the information available to the brain and on the perceived value assigned to that information. After many competing ideas are evaluated, they are prioritised – either by an individual, or a collective. The prioritised ideas are then willingly acted upon, through purposeful activity. Their implementation leads to change of the future states in comparison to that expected in the case of non-action. The purposeful activity eventually reshapes the future states, thus generating new information about reality. Through feedback between action, perception, new ideation, evaluation, and reprioritisation of ideas, the new information is perceived by the brain, assigned value, and recursively integrated into updated internal models, informing the generation and prioritisation of new ideas.

So, in this recursive closed-loop system of human cognition and action, ideas function as predictive hypotheses. They can prove more, or less, correct in the future. Their prioritisation acts as their selection mechanism. The following purposeful activity modifies reality. The outcomes generate new information for the brain to perceive and assign value. This updates brain's internal models, which can then generate, or perceive, new ideas. The described process is fundamentally compatible with Wiener's cybernetics [**7**], Ashby's adaptive systems [**8**], Friston's predictive processing and "free-energy" ideas [**9**], and "Bayesian brain" hypotheses [**10**].

We propose this process to be considered *a general theory of human progress*. To the best of our knowledge, there is not a widely accepted, unified "general theory of human progress". However, there are partial theories, each capturing an important dimension of human progress, but none of them integrating them all in a single testable framework which is compatible with the existing theories. The most directly relevant existing "partial theories" on which a general theory of human progress needs to be built, are very briefly summarised in **Table 1**.

**Table 1:** A brief summary of the existing "partial theories" of human progress, on which a general theory of human progress needs to be built.

| EXISTING (PARTIAL) THEORY | MAIN CONTRI-BUTORS | UNDERSTANDING OF HUMAN PROGRESS | STRENGTHS | WEAK-NESSES |
|---|---|---|---|---|
| Economic growth & innovation theories [**11**,**12**] | Adam Smith, Robert Solow, Paul Romer | Initially as accumulation of capital and technology, later refined into endogenous growth, where ideas drive growth, and with a key related insight that ideas are essentially non-rival, and can scale infinitely | Focus on macro-economic indicators | No insights into how ideas are generated, selected, and perceived |
| Evolutio-nary and cultural selection theories [**13**,**14**] | Charles Darwin, Richard Dawkins | Initial variation leads to natural selection and then to retention of genes in biology, and memes in culture, where both genes and memes compete within their specific contexts | Process of selection within multi-dimensional context | No definition of evaluation criteria, including conscious judgement |
| Philosophy of science, knowledge growth [**15**,**16**] | Karl Popper, Thomas Kuhn | Progress in science and knowledge growth comes from conjectures and refutations (Popper), or paradigm shifts (Kuhn) | Highly applicable to scientific ideas | Low relevance for non-scientific ideas, decision making |
| Enlighten-ment and human progress narratives [**17**,**18**,**19**] | Historically Voltaire, John Locke, Immanuel Kant; now Vaclav Smil, Steven Pinker | Progress is defined and measured through improvements in human health, wealth, safety, and knowledge; it is driven by reason, science, and strong human-built institutions | Quantitative basis in empirical trends | No insights into mechanisms that lead to progress or regress |
| Complexity and systems theories [**20**,**21**] | Joseph Tainter, Peter Turchin | Progress is understood as an emergent property of complex systems, and it is based on societal "cliodynamics" – mathematically modelled | Recognising and describing macro-patterns | No insights into micro-level idea selection |
| | | cycles, collapses, and feedback loops | | |
| Information theory [**22,23**] | Claude Shannon, Gregory Bateson | Progress is linked to better information processing, where increased knowledge reduces uncertainty | Appreciates the importance and value of information as central concept | Does not have decision and prioritisation layers for real-world application |
| Expected utility theory, Decision theory [**24,25**] | Nicolaus Bernoulli, John von Neumann, Oskar Morgenstern, Leonard J. Savage, Frank Ramsey | Progress is viewed as cumulative result of individuals and institutions making decisions that maximize expected utility under uncertainty; consistent choice of actions whose anticipated benefits, weighted by their probabilities, exceed the expected costs and risks | Rigorous mathematical framework for rational choice; integrating uncertainty, preferences, and probabilities into a single decision rule | Assume a fixed set of options and focus on selecting among them, do not consider a full "idea cycle" over space and time and how it innovates |

The seven theories shown in **Table 1** all bring useful perspectives to understanding human progress, but they all miss some of the important components of the ideometrics-based general theory of human progress: (i) generation of ideas (what is their point of origin?); (ii) evaluation of ideas (which criteria should "measure" the value of ideas in different contexts?); (iii) prioritisation of ideas (under constraints of human capacity, time, energy and resources, which ideas will be pursued and implemented?); (iv) the important role of consciousness, which enables brain's "sense of ideas" and "valuation of information"; and (v) quantifiability and testability in strictly scientific terms, using methods and tools that allow well-designed trials and refutability of hypotheses. The existing theories remained difficult to generalise in this way because they were either too field-specific (e.g., economic growth, biological evolution) [**11-13**], abstract (e.g., philosophy of science, cultural evolution) [**14-16**], descriptive rather than "mechanistic" (e.g., enlightenment, complexity and systems, information theory) [**17-23**], or focused primarily on rational choice under uncertainty (e.g., expected utility theory, decision theory) [**24,25**].

The framework that the ideometrics-based approach could bring into understanding of human progress is substantively different from previous theories: it offers a unified, "mechanistic", and potentially quantifiable theory of how ideas drive all human progress. In the proposed role of the brain as a "sensor of ideas", it was stated that "*…pursuing various ideas tends to drive most of human activity. It requires prioritising between short-term, mid-term and long-term investment of human energy and time*" [**1**].

Therefore, some of the "missing elements" that the ideometrics-based theory could bring into equation are: (i) human brain as a "sensor of ideas", thus connecting biology, consciousness, sense of time and space, and cognition; (ii) the concept of "value of information" assigned by the brain, thus connecting epistemology, memory, information processing, and decision-making processes; and (iii) the CHNRI method, as an example of a tool that links generation, evaluation and prioritisation of ideas, through a full "idea cycle". The method also brings together human collective knowledge and opinion using crowdsourcing and taps into the "collective sense of ideas". It assigns "value" to competing ideas based on criteria that best describe the context in which the prioritisation occurs. The quantitative nature of the method and its outputs allow for trials that can compare the outcomes in the presence and absence of this ideometrics process.

These "missing elements" may suggest some of the possible reasons why a general theory of human progress was not proposed in the past. It requires integrating many diverse fields of science through a trans-disciplinary approach, to create novel conceptual frameworks. It needs to bring together neuroscience, philosophy, economics, decision science, information theory, social sciences and humanities, cybernetics, artificial intelligence (AI), and possibly some other fields. Most of the existing theories are predominantly rooted in one area of science, while this theory is transdisciplinary – just like the CHNRI method itself, developed through a transdisciplinary collaboration to eventually give rise to the emerging field of ideometrics [**3**].

The existing theories of human progress mainly aim to explain what progress looks like. One of the key conceptual advances of the proposed ideometrics-based theory is its attempt to explain how human progress is, essentially, *being selected* through competition of ideas, and how the quality of that process can be observed and measured. This selection occurs continuously, within an observable context and through measurable and testable processes centred on ideas. The earliest roots of this novel framework can be found in the earlier description and analysis of the key conceptual challenges faced when setting priorities in health research investments is required [**26,27**].

**Definition of "human progress" in ideometrics terms**

To develop ideometrics into a recognised scientific field and provide credibility to a "general theory of human progress", a formal and testable definition of "human progress" will be required. In ideometrics terms, human progress is *"measurable improvement in the ability of individuals and societies to generate, evaluate, prioritise, and implement ideas – under scarcity of human capacity, energy, time and resources - in a way that increasingly aligns prioritised ideas with those that truly lead to preferred future states, given available information and uncertainty"*.

There are several components of this definition that require further thought. Firstly, it should be noted that ideas are the fundamental building blocks of this theory. As a result, "human progress" is not defined by increase in human wealth, health, or available technology, but rather by the ideas that were generated, evaluated, prioritised and implemented to eventually produce those outcomes. This also makes the well-functioning "sense of ideas" at the level of individuals and societies the key determinant of human progress.

Second, to predict human progress in the future, it is not that important what have humans achieved historically; it is considerably more important how good are they, as individuals and a collective, at choosing which ideas to pursue next. This avoids a shortcoming of the existing theories, where only outcomes are measured. It also explains why past outcomes often do not predict future outcomes well. The underlying reason for divergence between the past and the future can be found in a major change in the quality and efficiency of ideometrics systems on which the progress is based. This quality of the ideometrics processes is what needs to be measured in this new conceptual framework.

Third, the ideometrics perspective on understanding human progress considers a full ideometric cycle, acknowledging that progress can result from improvements in multiple domains: increasingly good ideas being proposed (generation), improved criteria for their assessment, which are more relevant to the context (evaluation), and improved institutions and processes that ensure choosing the "right" ideas (prioritisation) and acting upon them effectively (implementation). Importantly, failure in each one of those steps of the "idea cycle" can undermine, slow down or entirely break the progress, or even lead to regress.

The fourth component could be thought of as the "hard problem" of the ideometrics-based, general theory of human progress. That is the concept of "preferred future states". Clearly, something that is "preferred" by a conscious individual, or a group of humans who aligned their interests, does not need to eventually prove "better" even for themselves, let alone for many others. Furthermore, it is difficult to objectively physically measure what is "better" for anyone, and at which exact point in time should that be assessed. This is because, when people live long enough, they often realise that many of the life outcomes once perceived as positive turned out to be perceived negatively in the longer term, and vice versa – some past outcomes initially perceived as negative could have been "blessings in disguise".

This is where the ideometrics-based theory of human progress returns to the initial conceptual challenges, faced two decades ago by the transdisciplinary team of developers of the CHNRI method for setting health research priorities [**26**]. The same "hard problem" was undermining their efforts to develop a fair and legitimate process to prioritise between many proposed research ideas in the face of uncertainty. The passing of time can change the preference for many outcomes through a lifetime. Another layer of complexity is the subjectivity of what a "better future" means for individuals and human groups. The "preference" for the outcomes depends on human values, goals, and perceptions, many of which are diverse and rather elusive to objective testing. In the most general ideometrics framework, based on brain's "sense of ideas" [**1**], it was suggested that individuals tend to prioritise their ideas based on their personal emotional ("attractiveness"), rational ("feasibility"), and forecasting criteria ("predicted impact").

However, there may be at least some outcomes that most individuals and collectives "prefer" because they are rooted deeply enough in evolutionary and natural selection principles. Examples are long-term survival, good health, safety in daily life, and financial stability. These are all well-aligned with evolutionary drive towards survival of the individuals and entire species. Extinction has already affected more than 99.9% of species that ever inhabited the Earth, with an average lifespan of a species estimated to about 1 to 10 million years, but varying widely between taxa [**28**].

With at least some universally “preferred future states” for humans identified, we can provisionally define a “preferred future state” as one that is assigned higher expected “value” by a conscious observer – either an individual, or a collective. In ideometrics, it would be important to identify a set of observable, measurable and quantifiable “preferred future states” that could be compared to their alternatives mathematically and statistically. This brings us to the fifth component of the definition of “human progress” - its measurability. Without measurement, ideometrics would collapse into a philosophical construct. Because of this, the general theory of human progress based on ideometrics must be tested through measurable proxies of quality of idea generation, valuation, prioritisation and implementation, and through measurable outcomes, i.e., “future values” of prioritised ideas.

**Defining a measurable “Ideometric Index of Human Progress” (IIHP)**

To move the ideometrics-based view of human progress from purely conceptual to practically applicable, a measurable index of human progress will be required. This index would need to measure *the quality of idea selection processes under uncertainty*. Therefore, the *Ideometric Index of Human Progress* (IIHP) should be constructed as a composite, time-dependent measure of how effectively a complex system – whether an individual human, a private or public institution, or the whole society - selects and implements ideas that achieve higher realised future value relative to available alternatives. The index has three components, reflecting the full ideometrics framework.

*Component 1: Quality of ideometrics process (Qi)*

This element should evaluate the quality of the entire process of generating, evaluating and prioritising ideas, which can be based on the CHNRI method, or on some other ideometrics method [**4**]. The starting point here is scarcity of capacity, energy, time and resources that can be invested in pursuing all the possible ideas that could be envisaged. In practice, this problem is shared by those who decide on investment portfolio when considering all proposed stock market shares, start-up companies, or research grants – the latter being addressed by the CHNRI method over the past two decades [**3**]. The CHNRI method will give better results if the set of initially proposed ideas is comprehensive and sufficiently diverse (i.e., it does not miss many ideas of high real value at some future time $t$, $V(i,t)$). This part of the ideometrics process will be captured by the parameter $G$, which stands for idea generation quality. The next task for this theory will be to develop quantitative indicators of quality of generating ideas.

The process will also give better results if the criteria for evaluating ideas are chosen well. This will be true when they reflect the true nature and complexity of the context in which prioritisation takes place [**26**]. They need to capture a very large proportion of the inherent contextual variation, which may require placing appropriate weights to each criterion [**26**]. Also, the process of idea evaluation will be of higher quality if genuine experts are taking part in the process [**29**], and if they number is sufficiently large to allow for a replicable crowdsourcing [**30**], as we showed in our earlier work. Then, the quality of idea evaluation – which is central to ideometrics - will be captured by the parameter $E$, which stands for idea evaluation quality. It measures how well is the “perceived future value” of many proposed ideas – expressed quantitatively using the chosen criteria and collective expert judgement, or using AI - grounded in the best possible evidence and information at the present time.

This ideometrics process will also be of higher quality if those ideas that achieved the highest priority scores through the evaluation stage of the process are eventually prioritised and selected for implementation, rather than dropped because of some undue influences. This will be captured by the parameter *P*, which stands for idea prioritisation efficiency. This parameter should be easily quantifiable, because it will be strongly associated with the ranks that the ideas achieved in the previous step of evaluation.

Then, we can express the quality of the ideometrics process as:

$$\mathbf{Qi = G \times E \times P}$$

where it can be computed as a *product of idea generation quality, idea evaluation accuracy, and idea prioritisation efficiency*.

*Component 2: Effectiveness of implementation*

In addition to those three steps of a typical CHNRI process, we need to recognise that even the ideas with the greatest future real value will eventually fail without their effective implementation. Therefore, in this element of the index, we will need to quantify the effectiveness of implementation of the prioritised ideas (*Ie*). The simplest possible measure would be:

$$\mathbf{Ie = I(t) / S(t_0)}$$

This simplistic measure of the effectiveness of implementation (*Ie*) is a fraction, with the number of selected ideas that are successfully executed at some future time *t* (*I(t)*) in the numerator, and the set of ideas that was prioritised at the time of completion of the ideometrics exercise ($S(t_0)$) in the denominator. However, this simple form doesn’t really capture the essence of this part of the “idea cycle”. It will require further improvement, because it needs to measure the capacity of individuals and societies to effectively implement all the prioritised ideas. Therefore, this is not only about how many of those prioritised ideas were eventually implemented, but also how skilfully and effectively did the implemented ideas change the reality towards preferred states. So, this component of the IIHP will require more thought.

Now, putting all these elements together, we can express the Ideometric Index of human progress (IIHP) as:

$$\mathbf{IIHP = (G \times E \times P) \times Ie = Qi \times Ie}$$

If we can measure this index in reality using publicly available records over long periods of time and base the measurements on well-chosen examples (e.g., stock-market investing, investing in start-up companies, national-level funding of research grants), and base computations on sufficiently large and comprehensive datasets, ideometrics could evolve into practically implementable and testable field of science.

**Measuring the "Ideometric Index of Human Progress" (IIHP) in reality**

Clearly, the next challenge is finding real-life examples and approximators of the values *G*, *E*, *P*, and *Ie*. Based on the long-term experience with implementation of the CHNRI method, we will provide some early suggestions, although better approaches may emerge over time:

*(i) G = Idea Generation Quality*

During the development of the CHNRI method, one of the key challenges that the transdisciplinary team addressed was how to reassure the users in completeness of the initial list of ideas. The expectation was that most of the potentially valuable ideas would be included in the initial set. Our approach firstly defined the "4D framework" that was relevant to health research (i.e., "description", "delivery", "development" and "discovery" instruments of research) [**26,27**]. In a more general application, this would correspond to: (i) "ideas how to describe the challenge"; (ii) "ideas how to improve the efficiency of addressing the challenge using the existing resources"; (iii) "ideas how to improve the existing resources themselves"; and (iv) "ideas how to develop entirely new, and better, resources to address the challenge". Here, the "challenge" can be literally anything - from improving personal health or finances, to major societal challenges such as violence, inequity, climate change, or weak economic growth. Proposing many ideas that belong to each of these four general "instruments" of progress secures a higher degree of completeness.

As the implementation of the CHNRI method expanded [**32**], we observed that with each new idea proposed to the initial set, the probability that a similar idea was already contained in the set started to increase. This was particularly apparent when subsets of 10 or 30 successive ideas were analysed in comparison to all those previously received. This implied that, although the set of possible ideas at the start of the ideometrics process is theoretically infinite, there is a detectable saturation in practice. For each practical exercise that we conducted, a suggestion of saturation point, beyond which there was little additional novelty, seemed to genuinely appear. Therefore, the initial ideation leads to a point where majority of ideas that are practically feasible at present have been collected, and it would take many additional ideas to find some truly new in comparison to the rest of the set.

This finding has interesting implications in the age of AI-based large language models (LLMs): large-scale AI-based idea generation and AI-based text analyses are both becoming feasible. This should allow an empirical generation of very large number of ideas on how to address a challenge, followed by a statistical approach based on text processing to develop measures of diversity of idea sets and progress of idea saturation in each set. It may be possible to even develop measures of semantic distance between individual ideas, and measures of novelty of individual ideas based on their semantic distance from the existing knowledge base, as well as the diversity within the entire initially proposed set of ideas.

*(ii) E = Idea Evaluation Accuracy*

Accurately evaluating the future real value of ideas, by approximating it with perceived value of ideas based on the best possible evidence and information available in the present, is the core of ideometrics. The measure *E* is about ensuring the quality of that process in the present, because the actual alignment between predicted and realised *V (i,t)* can only be measured "post hoc" at some time *t* in the future. To this end, we could compare predicted rankings of the ideas in, e.g., several historic CHNRI exercises, with later frequency of prioritised research ideas in the literature; or analyse the structure of past stock market portfolios and their performance over different points in time.

However, the main issue related to measuring parameter *E* in the present is about developing a quality system for evaluating ideas that should be reasonably accurate in predicting the future realised *V(i,t)* at different time points. Interestingly, many different disciplines tend to converge to a similar framework when trying to predict the future value of ideas, and they are based on multiple criteria that can assign a greater likelihood of success to one idea over the other. The reason for using several different, ideally non-correlated criteria, is to try to capture multi-dimensional nature of the context in which the value of the ideas will eventually need to be realised. This context is often highly complex and unpredictable, but some of its key elements can be identified and captured in the chosen criteria. The challenge, then, becomes to "measure" how each proposed idea satisfies each criterion. This can be done using crowdsourced expert opinion, as the CHNRI method demonstrated [**26,27**].

However, a deeper issue is to reduce uncertainty over: (i) how many criteria are needed to capture most of the complexity of the context that will affect realisation of idea's value in the future; (ii) which exact criteria to choose; and (iii) how to address the fact that not all criteria should carry the same "weight", because some are more important than others? In statistics, there are analogies between this problem and Principal Component Analysis (PCA), which explains the variation observed within a dataset by identifying orthogonal combinations of variables that explain the largest proportions of variance [**33**]. So, if there is an opportunity to measure the variation of the relevant characteristics of the context in which ideas will be implemented, then a PCA - if at all feasible - could help suggest the most useful criteria.

There may be a deeper, more fundamental reason why this kind of approach is useful. Across many scientific disciplines, selection processes tend to favour innovations that perform sufficiently well across multiple relevant dimensions of their context, rather than those that excel in only a single dimension while failing in others. This phenomenon can be observed in systems ranging from novel genetic mutations undergoing natural selection, to memes spreading on the internet, global university ranking systems, research grant evaluations, and start-up company selection. In all these cases, the probability of selection appears to depend on the aggregate performance of competing innovations across multiple contextual criteria within the environment in which competition occurs. This suggests that the criteria used in ideometric evaluations should aim to approximate the principal dimensions that jointly determine the probability of successful selection and long-term impact.

Once the criteria are chosen and the weights are assigned to capture the context in which idea evaluation occurs, our previous work showed that the process of idea evaluation will be of higher quality if genuine experts are taking part in the process [**29**], and if they number is sufficiently large to allow for a replicable crowdsourcing [**30**]. Therefore, a quantitative index will be required to capture the quality of an ideometric method that evaluates the ideas. It will need to assess whether all reasonable measures took place to choose the most informative criteria given the context, to assign them appropriate weights, and then that the most knowledgeable experts were invited to the process and that their number was large enough to ensure replicability. The question then remains whether AI, which has been trained on the entirety of human knowledge, could also perform this task, which will lead to comparative studies of the outcomes of expert crowdsourcing and AI, which we already conducted [**31,32**]. If we demonstrate that AI and humans can indeed be used interchangeably, then this framework based on scores against multiple weighted criteria will serve to make the AI-based evaluation of large number of ideas explainable to humans, which should also be useful.

*(3) P = Idea Prioritisation Efficiency*

In theory, it should be simple to prioritise and select the ideas that received the highest scores through evaluation. In practice, however, the likelihood of selecting the ideas with the highest scores will be related to the incentives of those who hold the governance-related powers over priority-setting exercises and stand to either benefit, or lose, from the consequences of an ideally efficient, i.e., "meritocratic" implementation. Therefore, *P* becomes a measure of "decision discipline": it can capture the influence of politics, various biases, possible corruption, institutional inertia or incompetence, and similar undue influences.

However, there are some opportunities that the recent CHNRI processes provide. Methodological advances made it possible to quantify uncertainty of the achieved research priority scores, statistically evaluate inter-scorer agreements, and search for evidence of bias and unexpected groupings of the scorers. Those solutions have already been incorporated in the most recent CHNRI exercises, and they assist the process of prioritisation after the completion of evaluation [**31,32,34**].

As mentioned earlier, this parameter should be more easily quantifiable because it is closely associated with the ranks that the ideas achieved in the previous step of evaluation (*E*). Hence, the smaller the average rank of all the ideas that are prioritised and implemented, the better the prioritisation element has worked.

But even in an ideal case, with perfectly valid scores, hardly any uncertainty about the rankings, supportive agreement statistics and no detectable biases, there is still an important remaining need in real world scenarios: how to specifically allocate funding to individual prioritised ideas from a fixed budget? Thus, there may still be an additional layer required here, to optimise the prioritised portfolio of ideas taking in account possible positive and negative interactions among the ideas that were prioritised. This is a problem of investment portfolio optimisation that will need to be resolved methodologically to make the CHNRI process even more helpful to investors. That aside, the small average rank of prioritised ideas would be an indicator for a well-performed prioritisation part of the ideometrics process.

*(4) Ie = Idea Implementation Effectiveness*

This part of the "idea cycle" is currently underdeveloped in a typical CHNRI process, and it will require additional work. Clearly, comparing "ideas implemented as intended" and "ideas prioritised for implementation" will require some adjustments for delays, fidelity to original aims, cost overruns, etc. Our group and our external collaborators tried to address some of these needs by developing the "Planning, Monitoring and Evaluation Tool" (PLANET) [**35**]. Further work in this space should improve measurement of implementation success for prioritised ideas in real-world scenarios.

**Interpreting the value of the Ideometric Index of Human Progress (IIHP)**

Let us now consider what values might IIHP acquire in real world measurements. If we ensure the quality of idea generation through use of AI-based Large Language Models (LLMs) and measure saturation, diversity and novelty within the initial set of proposed ideas, it is possible that this component of the index could acquire quite high values, say up to 70-80% of theoretically maximum completeness. Then, let us assume that it is also possible to measure variation in many variables relevant to the context with a great degree of precision, and then use PCA to identify several most useful criteria that could jointly explain about 70-80% of that variation. This will likely be further reduced when measurement uncertainty and the imperfections of the process of expert scoring are taken into account. So, let us assume that the quality of evaluation of the ideas in real world scenarios may realistically reach about 50-60% of the maximum possible.

Are those estimate realistic? If those proportions were much higher, then it would be easier to predict future outcomes accurately. In principle, it is not impossible to predict even longer-term future very accurately, under certain conditions. As an example, we could predict the position of the Moon in the night sky exactly 300 years from now with a remarkable degree of precision, provided that no large objects hit the Moon in the meantime, and that no impact from human activity changes its trajectory over that period. In this case, an accurate prediction of the future is possible, because the system in question is simple enough and governed by laws of physics. There are not many variables that could significantly affect the outcome of interest. But in real world scenarios on Earth, the number of variables that can significantly affect the outcomes is much larger, systems and contexts are far more complex, so it is frequently not possible to make future predictions with such precision.

The next two parameters, prioritisation efficiency *P* and implementation effectiveness *Ie*, could both reach quite high values in realistic scenarios wherever well-minded, enthusiastic teams oversee implementation of ideometric processes. Let us assume that, with the development of good budgeting and portfolio optimisation methods, followed by an enthusiastic implementation, the values of both *P* and *Ie* could reach as high as 90%. Entering all those values in the proposed formula, IIHP = 0.75 x 0.55 x 0.90 x 0.90 = 0.33. Perhaps it could also be measured as an unweighted or weighted mean of the four components, in which case it would have higher values. However, the multiplicative nature of the IIHP proposed here serves the purpose to show that failing in any of the elements of the "idea cycle" will inevitably affect the entire progress.

Therefore, whenever d/dt(IIHP)>0, i.e. when IIHP is increasing over time, this will act as an enabling mechanism for "human progress", provided that the predicted future values of prioritised ideas were broadly correctly assessed during the evaluation stage (*E*). Other measures of progress, like Gross Domestic Product (GDP) or Human Development Index (HDI), are observational and "static" in nature and cannot easily predict future outcomes. However, the IIHP measures the quality of the process that generates those outcomes, and not the outcomes themselves, so it can be predictive of future progress and regress, given that the evaluation stage (*E*) is successful in predictions of future values for assessed ideas.

**Monitoring Future Outcomes (O)**

It is important to note that all the components of IIHP are measured in the present time and aligned with the operationalisation of each step: generating, evaluating, prioritising and implementing ideas. So, IIHP can be a useful indicator of how well the process of selecting "best ideas" functions in any society. It can signal if there are efforts in place to enable continuing selection of the "best ideas", or whether the process might be deeply flawed in one or several stages. Therefore, IIHP could be useful for raising the alarm that the ideometric process is compromised in a society.

However, it is very important to note here that, although well-minded and highly skilled policymakers may indeed ensure that the parameters *G*, *P* and *Ie* are maximised within their context, and that all the requirements for the high parameter *E* are in place, all of this may still not be enough to secure "human progress". Whether the progress occurs, and at what rate, will eventually depend entirely on how correctly did the invited experts, or AI, predict the future value of assessed ideas, and whether the ranking of ideas for prioritisation truly clustered the ideas with the highest true future value near the top of the ranks. However, this can only be assessed at various points the future, by monitoring the outcomes of implementation of the prioritised ideas and comparing them to the predicted outcomes. The outcome of the ideometric process (*O*) at some future point in time (*t*) can be expressed as:

$$O = \sum (\boldsymbol{V}(\boldsymbol{i},\boldsymbol{t}) \mid \boldsymbol{i} \in \boldsymbol{I}(\boldsymbol{t}))$$

This simply means that the outcome value can be computed as a sum of realised future value $V(i,t)$ of all implemented ideas that were eventually implemented, i.e., that belonged to the set *I(t)*. The practical value for *O* may be quite easy to measure, e.g., when firstly picking stocks on the stock market based on their perceived future value, and then checking the actual value at some later points in time. For those start-up companies in which investments were made, their future value can be measured as their annual earnings or profits at different future time points. For research grants that were chosen for funding, the number of papers and patents and their citations can be monitored. At the national level, results of policies that were supported in the parliament can be assessed and the trends in GDP followed. Even at the personal level, the chosen ideas can be written down in the present, and their effects then assessed in the future. In principle, it should be possible to find many examples where ideometrics-based predictions of the future values of prioritised ideas can be tested – both prospectively and retrospectively.

However, even after monitoring and measuring the outcomes quite precisely, as in the above examples, a question remains: how can we know if the observed future value (*O*) truly represents "progress", and how large has this progress been over that period of time in comparison to some alternative prioritisation of ideas, that could have been implemented instead?

**Understanding of "human progress" (HP) in ideometrics terms**

From previous sections of this paper, we can assess the quality of the ideometric process using IIHP, and then measure the outcomes of that process at some future point, as *O*. But what if the ideometric process had some systematic or fundamental flaw – e.g., the criteria for evaluation of ideas were not chosen well, or the experts who evaluated ideas have been deliberately misinformed? Then, even a rigorously conducted ideometric process, followed by excellent implementation, could still pick wrong ideas and lead to poor value and regress, instead of progress. We need to use the described framework to explain how we can think about human progress in terms of ideometrics.

Let us define *HP* as "human progress", and *I* as set of all theoretically possible ideas that could be chosen to contribute to *HP*. The set *I* is analogous to large sets of ideas that are initially proposed in the first step of any CHNRI exercise, or all stocks available for investing in stock markets, or all start-up companies that exist within a geographic area, or all research grant proposals submitted to a funding call. Then, let us define *V(i,t)* as a true "value" of a single proposed idea *i* at some specific point in the future *t*. That future value *V(i,t)* will materialise over time, so it is true, but it remains unknown to us. Then, let us define $\hat{V}(i,t)$ as a perceived future value, in present time, of a single proposed idea *i* at some specific point in the future *t*, based on the best available evidence and information that an individual or a group of individuals possess. Let us also define $S(t_0)$ as a set of prioritised ideas at time $t_0$ when they were chosen to be acted upon.

Then, whether the "human progress" occurs or not, following the ideometric process, will depend largely on the alignment between those ideas that were prioritised, and the ideas that will genuinely have the greatest value at time *t*:

$$\mathbf{HP}(\boldsymbol{t}) \propto \mathbb{E}[\boldsymbol{V}(\boldsymbol{i},\boldsymbol{t}) \mid \boldsymbol{i} \in \boldsymbol{S}(\boldsymbol{t}\mathbf{0})] - \mathbb{E}[\boldsymbol{V}(\boldsymbol{i},\boldsymbol{t}) \mid \boldsymbol{i} \in \boldsymbol{I}]$$

Now, let us carefully consider what this formula implies. HP*(t)*, i.e., human progress at some specific future time point (*t*), will be proportional (∝) to the difference between probabilistic expectation ($\mathbb{E}$) of the true future value of the random variable *V(i,t)* under the condition that it has been prioritised for implementation at the present time (denoted as $i \in S(t0)$), and the value of the same random variable *V(i,t)*, but this time under the condition that it belongs to the set of all possible ideas that could have been chosen to contribute to human progress (denoted as $i \in I$).

In this case, the probabilistic expectation ($\mathbb{E}$) measures the average true future value of ideas in both sets, $S(t_0)$ and $I$. It subtracts the average value of all possible ideas from the average value of all ideas that have been prioritised. Therefore, if the average value of prioritised ideas at the future time point $t$ was, say, 10, and the average value of all possible ideas at the same point in time was 3, then the difference is 7. So, it is positive (>0), and human progress will be proportional to this value. However, if a human society managed to prioritise a subset of all possible ideas with the average value of 2, then 2-3 = -1, and the negative value on the right side of this equation implies that regress should be expected, instead of progress. That "regress" may still not mean that things will get worse, if all ideas in the initial set were progressive, or at least neutral to the progress. But the latter scenario will mean that it would have been better to pick ideas randomly, leading to "human progress in absence of ideometrics" (HPAI), than to use ideometrics and select the ideas with lower real future value than prioritisation at random would have achieved.

This explains, through ideometrics lens, that human progress in the future will be the fastest and the greatest if we aim to maximise the positive difference between the average real future value of prioritised and implemented ideas, and the average real future value of all possible ideas. Also, we must avoid a negative difference between the average real future value of prioritised and implemented ideas, and the average real future value of all possible ideas, as this framework explains that this would lead to regress. This "regress" may not be apparent in absolute terms, if all the ideas in the initial set are progressive in nature, but it can still be a "relative regress", given that the outcomes will be worse than if there were no ideometric processes in place, and the ideas were selected at random.

A more rigorous way, from the lens of economics, to understand the ideometrics-based theory of human progress is through the framework of expected utility theory and decision theory [**24,25,36**]. From a small set of natural and intuitively reasonable requirements on choice, it derives a powerful conclusion: a rational agent must behave as if they have a utility function over outcomes, a probability distribution over states of the world, and they choose actions that maximise expected utility. This framework reduces rational decision-making to two fundamental epistemic tasks. First, one must determine how valuable different outcomes are. That is, one must form judgments about what constitutes e.g. human flourishing, welfare, prosperity, freedom, knowledge, health, sustainability, or other dimensions of value. Second, one must determine what the world is like.

Within this framework, an idea can be thought of as a transformation from possible states of the world to possible outcomes. The future, however, is uncertain. The consequences of implementing an idea depend on the true state of the world: for example, on biological, physical, economic, social, or technological conditions that may not yet be fully understood. Different ideas generate different distributions over outcomes depending on what the true underlying state of the world really is. Therefore, more formally, an idea is a map

$$i: S \to O$$

Where S is a set of states of the world, and O is a set of outcomes. Let x denote the true state of the world. Then the true value of an idea at time t is

$$V(i,t) = u(i(x),t)$$

A rational decision-maker's estimate of V(i, t) is

$$\hat{V}(i,t) = \int \hat{u}(i(y),t)dp$$

where $p$ is a probability distribution that reflects the decision-maker's uncertainty over states of the world, and $\hat{u}$ represents the decision-maker's best estimate of the true utility function $u$. The decision-maker chooses to prioritise those ideas that maximise $\hat{V}(i,t)$. A sufficient condition for human progress towards desired outcomes is then:

$$\boldsymbol{\frac{d}{dt} \mid \hat{V}(i,t) - V(i,t) \mid < 0}$$

for all ideas *i*.

Consequently, when alignment between the perceived future value of ideas $(\widehat{V}(i,t))$ and their true future value *(V(i,t))* increases over time, prioritised ideas increasingly correlate with those with the greatest true future value. Thus, the derivative of the absolute estimation error, expressed as $\mid \hat{V}(i,t) - V(i,t) \mid$, becomes negative. Such a trend leads to future human progress, because the ideas that are being prioritised over time increasingly align with the ideas that have the greatest true future value. However, if the ideas that are being prioritised start to diverge from the ideas that carry the greatest future value, then the result above will become greater than zero, leading humans to regress. Therefore, in ideometrics, one of the central aims is to improve our ability to estimate the real future value of ideas. This formula states that the field succeeds when the correlation between estimated and true future value systematically improves over time.

We believe that this equation captures one of the most profound aims of both science and cognition: *to make our beliefs about the world progressively more consistent with reality*. Science, markets, democracy, and artificial intelligence (AI) all function better when estimated values converge toward true values. Therefore, over time, both humanity and AI should become increasingly accurate in distinguishing truly valuable ideas from less valuable ones. This definition differs from the existing ones because it unifies several areas (i.e., economics, decision science, information theory, policy, culture, and others). It explains both progress and regress, and it naturally explains the threat of misinformation for human populations, as it leads to misalignment between the prioritised ideas and the most valuable ones – the issue that we discussed previously [**2**]. This definition also incorporates the value of expertise in better estimation of *V(i,t)*, which our group and collaborators demonstrated empirically in our experiments on quantitative properties of human collective knowledge [**29**] and opinion [**30**]. Surprisingly, we found no similar attempts in the earlier literature to address these important questions [**29,30**].

This definition also incorporates the opportunity to compare the choices made by the artificial intelligence, as a new "idea evaluation engine" trained on human collective knowledge, with human collective expert opinion gathered through CHNRI exercises. The first comparisons of the collective estimation of $\hat{V}(i,t)$ based on expert opinion using the CHNRI exercise, and the estimation based on the commercial AI chatbots, have been published recently by our group and collaborators in two large CHNRI collaborations [**31,32**].

The usefulness of this definition can be demonstrated through allowing statements like: "*this society may have become well-developed economically, based on their past idea prioritisation practices; however, it will regress because their accuracy of prioritising ideas with the greatest future value is changing for the worse*"; or "*this small group has developed a superb methodology for accurately predicting future value of stocks in the market; so, their present 'investment ideas' for picking stocks are well aligned with the companies that later show the greatest increase in value".* In fact, stocks on the stock market are a good example of the key ideometrics principles: "preferred" universally means "higher future price of the stock", and their "unknowable" true value can be followed up daily into the future and measured precisely in USD.

**Extending IIHP to "civilisational progress" on Earth and in Space**

The next interesting question is whether this ideometrics-based general theory of human progress could be extended over the course of history to entire civilisations? Could an ideometrics framework explain how humanity selects among the competing ideas and thereby determines the trajectory of civilisations themselves? In ideometric terms, "civilisational progress" could be defined as the long-term advancement of a human civilization through generating, evaluating, prioritising and implementing ideas that improve knowledge, health, wealth, prosperity, sustainability, the capacity to shape preferred future states, and other human progress-related indicators.

In this framework, the ideas are the fundamental units of progress, the brain's sense of ideas perceives them and assigns them their future value based on the "value of information" concept, and the entire societies become large-scale idea selection mechanisms. The most important addition for measuring "civilisational progress" is the role of the institutions in encoding, documenting, preserving and transferring the accumulated evaluations of past ideas from generation to generation. Understanding the laws that govern this entire process could lead to scientific explanations of why societies prosper, stagnate, or regress.

There is no real limit to this framework in time or space. If it could be scientifically understood how ideas arise, compete, and self-organize to determine the long-term trajectories of individuals, societies, or entire civilizations, then those laws could potentially be applied even to other hypothetical intelligent life forms in the universe, thus connecting ideometrics even to a hypothetical astrobiology. A mathematical expression that could describe the rise, flourishing, stagnation, regression and eventual collapse of civilizations, as functions of their ability to generate, evaluate, prioritise, implement, assess, document and successfully transfer among generations highest-value ideas – but also, importantly, the worst ideas historically - would connect ideometrics with history. Thus, the central hypothesis would be that the fate of civilizations will be determined by the efficiency with which it performs the "ideometrics cycle" of ideas, with an addition of documentation and transmission of ideas. They need to retain successful innovations both institutionally and culturally but also suppress harmful ideas. When these processes fail, civilisations will start to decline.

One might imagine a formal expression of ideometric index of civilisational progress (IICP), such as:

$$\mathbf{IICP(t) \propto \int G(t) \times E(t) \times P(t) \times Ie(t) \times O'(t) \times D(t) \times T(t)\, dt}$$

where *G* is, again, generation quality of ideas, *E* is their evaluation accuracy, *P* is their prioritization efficiency, *Ie* is effectiveness of their implementation, *O*' is an appropriately transformed form of the outcome value that would fit with other parameters in this formula, *D* = a measure of success in documenting ideas with large value (but also those regressive and destructive), and *T* = success in transmission of acquired understanding of the value of ideas. This equation extends the IIHP's concept to societies and epochs, making history interpretable as a long experiment in idea selection.

The proposed formula indicates that "civilizational progress" will be proportional to the integral over time of seven multiplicative components, leading to "civilizational advancement" - a broad measure of cumulative societal development at time *t*, potentially reflected in combinations of life expectancy, scientific knowledge, technological sophistication, economic prosperity, institutional quality, environmental sustainability, and human well-being. The proposed IICP is, simply, a large-scale extension of the IIHP, where the variables from the original formula acquire slightly different meanings: *G(t)* becomes a measurable rate at which a society produces novel candidate ideas at time *t*, such as scientific hypotheses, technologies, policy proposals, institutions, artistic and philosophical innovations; clearly, a society with strong education system, many creative minds, and intellectual freedom will tend to have higher values of *G*. *E(t)* still measures how accurately the society estimates the true future value of competing ideas, e.g. through developed scientific institutions, peer review, CHNRI-like processes, other ideometrics methods, forecasting through prediction markets, democratic deliberation processes, and AI. *P(t)*, then, measures the extent to which the most valuable ideas are chosen for investment and implementation. Importantly, societies may generate and evaluate ideas well, but fail to act on them because of politics, corruption, or inertia.

Then, *Ie* is the society's ability to transform selected ideas into reality, e.g., build infrastructure, conduct research, deploy technologies, enforce policy. *O'(t)* still needs to be defined and transformed appropriately to capture outcome value of the ideas that were implemented within this framework. Two new variables are also added: *D(t)* measures how effectively successful ideas are documented and preserved, e.g., from cave drawings in early humans, to stone encryptions and hieroglyphs, libraries of the ancient world, to searchable digital repositories today. Finally, intergenerational transmission *T*(*t)* is key to pass the documented knowledge across generations. This is the purpose of the education system, implemented in early life through public and private institutions. According to ideometrics, better educated humans will be more aware of the context, which should assist them to priorities ideas of higher future real value.

The seven variables are multiplied in this formula, because all are necessary. If any component is near zero, civilisational progress will be severely constrained. Thus, the formula reflects a principle that “chain is only as strong as its weakest link” and exposes fragility of civilisations and routes to collapse. The integral in the formula reflects the fact that progress accumulates continuously over time. Each historical moment contributes an increment, while total progress of civilization is the sum of these increments from the beginning of its history to the present time.

A follow-up to this work could demonstrate that, e.g., the Scientific Revolution greatly increased *G* and *E*, digital storage expanded *D*, while modern education expanded *T*. In recent years, AI use may amplify all components simultaneously. Such analyses could show that human history can also be understood through ideometrics, and that ideometrics can be successfully linked with history. Importantly, this ideometrics view shows why history "repeats itself": whenever humans fail to retain the quality of "documentation” and “transmission", damaging ideas will likely resurface, because there will be diminishing capacity to remember their historic outcomes, so they will be able to rise again in the next generations.

Paul Romer emphasized that ideas drive economic growth [**37**]. This equation generalises that insight, by modelling the entire pipeline through which ideas are generated, evaluated, prioritized, implemented, assessed, documented and transmitted between generations. The equation suggests that history is not a random process, but it can be understood through ideometrics and its derived models. Civilizations will rise when they become increasingly effective at discovering and acting upon valuable ideas. Their progress will then be dependent on the time-integrated consequence of collective idea selection. This makes civilizations vast ideometric engines that search idea space and convert successful ideas into their enduring reality and progress, or regress and collapse.

The framework could even be extended to a hypothetical astrobiology, by linking to the Drake Equation [**38**]. The probability that intelligent life persists may depend on its ideometric efficiency, i.e. its capacity to identify and implement beneficial ideas while avoiding destructive ones. This could also connect Ideometrics with the Fermi Paradox [**39**]. In this view, “survivability” of an alien civilisation would be associated with its capacity to navigate the space of possibilities in time and space through ideas, and “civilization” would be the collective embodiment of this capacity. At the level of space, “civilisational progress” would be the cumulative movement toward increasingly preferred future states.

**Current weaknesses of the ideometrics-based theory of “human progress”**

The weaknesses of this view, which need to be acknowledged, are the following: (i) true value $V(i,t)$ remains unknowable, and we can never fully know or observe it in the present; we can eventually observe the outcomes, but unlike the values of the stocks in the stock market, other outcomes important to individuals and societies will often be delayed and noisy, confounded by the outcomes of many other ideas that will be implemented and followed in parallel; (ii) preference is a subjective category, so “preferred future states” will differ across the ones perceiving those ideas; as explained, preferences will be affected by cultural conflicts, politics and ideology; (iii) the problem of time horizon is present again here, just as it was when the CHNRI exercise was being developed two decades ago [**26**]; it reopens the issue of short-term vs long-term value misalignment, and that some ideas may lead to preferred outcomes in short term, but unwanted outcomes in long term; (iv) there is also "path dependency", because once selected and acted upon, the ideas will reshape the future context, which then might lead to a need of re-prioritisation of the selected "portfolio" of ideas. In any case, this definition implies that human progress is not about moving toward a preferred goal, but rather about *improving the process that chooses its preferred goals*. It is mediated by the brain's perception of ideas [**1**] and the value assigned to information [**2**]. Its quality can be assessed by IIHP, while its effects on the human progress can be approximated by measuring the alignment between perceived and realised outcomes of selected ideas over time.

**Discussion**

This paper attempts to provide a transdisciplinary, formal, and potentially testable general theory of human progress based on the generation, evaluation, prioritisation, implementation, and validation of the outcome of ideas – i.e., through an "ideometric" approach. Thereby, "ideometrics" can be thought of as an adaptive control system that minimises the discrepancy between predicted and realised value of proposed ideas through recursive feedback. This makes the theory aligned with the well-known principles of cybernetics, Bayesian inference, and predictive processing [**40**].

By increasing the likelihood of "preferred future states", humans may reduce uncertainty regarding all their possible futures, minimising their collective informational entropy. This may represent one important functional role of consciousness – to allow the brain's "sense of ideas" to prioritise ideas that are more likely to secure safety, health, survival, and in modern times, financial stability and access to a wider range of future opportunities [**41**]. A valid question could be posed: whether "preferred future states" necessarily imply "human progress", especially when preferences differ radically among individuals and groups? The ideometrics framework proposed here is, in fact, agnostic to the content of preferences. It mainly focuses on the quality, accuracy, efficiency and effectiveness with which ideas are prioritised and implemented.

Some "hard problems" of ideometrics-based general theory of human progress, which were also inherent to CHNRI development, still remain: (i) estimating "true value" of a present idea in the future, possibly through analysing proxies from many historic series in different areas of human activity (e.g., DALYs averted for health; or financial and economic returns; or scientific impact, patents and case studies); (ii) subjectivity in values that different stakeholders will assign to criteria for idea evaluation; (iii) addressing the effect of time lag for evaluating outcomes; (iv) predicting how the time lag would change the future context, and if re-prioritisation of ideas should be required in regular intervals (this was one of the early suggestions of the CHNRI method – to mitigate this problem by conducting re-prioritisation in regular time intervals, or explore approaches that use rolling cohorts and predictive validation, with possible examples of research grants, or spin-out companies); (v) counterfactual problem of knowing what would have happened if those ideas that were not financially supported received equal support, and are there any "natural experiments" of this. It may also be possible to use simulations, historical comparisons, synthetic data and AI to model this.

The novelty of the theory proposed in this paper lies in a specific conceptualisation that, to our knowledge, does not yet exist as a coherent and testable framework. It builds on several well-established concepts from different fields of science: (i) that ideas drive human progress, which is well established in economics, e.g., through Romer's endogenous growth [**37**]; (ii) that selection processes among competing innovations matter, which is known from Darwin's evolutionary theory and Dawkins' memetics [**13,14**]; (iii) that information shapes decisions, which is known from Shannon's information theory and Bayesian reasoning [**10,22,23**]; (iv) that collective decision-making works, as demonstrated by e.g. Ostrom in her work on governance of common-pool resources, showing that decentralised groups can self-organize and make effective collective decisions [**42**], and later through the application of crowdsourcing, prediction markets, and the "wisdom of crowds" concept, as summarised by e.g. Surowiecki and others [**4,43**]; (v) that ideation for decision-making can be quantified, which I summarised recently with Sheikh (e.g., the CHNRI method, multi-criteria decision analysis (MCDA), cost-effectiveness, etc.) [**4,34**]; (vi) that human psychology and economics are connected, as demonstrated by e.g. Thaler [**44**], and that systematic biases in human judgement exist, as shown by Kahneman [**6**]; (vii) that decision-making is inherently constrained by limited information and computational capacity, as formalised by Simon through bounded rationality and satisficing [**45**], and that rational agents can nevertheless make robust choices under uncertainty and model misspecification, as demonstrated by Hansen and Sargent through their work on econometric inference and decision-making under model uncertainty [**46**]. The present theory synthesises these previously independent insights into a unified framework in which ideas are treated as quantifiable entities that can be generated, evaluated, prioritised, and implemented through both individual and collective processes, thereby seeks to unify these earlier perspectives into a broader science of ideas themselves.

The novelty of the ideometrics-based general theory of human progress lies in proposing a single, causal framework, simultaneously integrating generation, evaluation, prioritisation, implementation, and outcome assessment of ideas, as well as their documentation and transmission, into a single, closed, dynamic and adaptive system. Also, it makes a shift from measuring outcomes to measuring decision processes, because most frameworks measure the outcomes such as GDP, health, scientific output, or company growth. Our theory proposes to measure the quality of the process of selecting ideas that produce those outcomes, which is a different focus of analysis. There is also formalisation that associates "human progress" with high quality of ideometrics processes coupled with increasing alignment between perceived and realised future value of ideas.

The theory also manages to bridge cognition and consciousness with societal systems and outcomes, starting from the brain's role as "sensor of ideas", individual evaluation of ideas, collective scoring (e.g., through the CHNRI method or other ideometric processes), and societal outcomes. Therefore, the theory spans from biological neurons to implemented global policy. Its IIHP and IICP indices, unlike GDP, HDI and other familiar measures of human progress, attempt to quantify how well a system chooses among competing ideas under uncertainty.

Paul Romer's work on endogenous growth theory [**37**], and expected utility theory and decision theory introduced by Bernoulli, Von Neumann, Morgenstern, Savage and Ramsay [**24**,**25**], are perhaps the closest "relatives" in the field of economics to the ideometrics-based general theory of human progress proposed here. Romer showed that ideas are central drivers of long-term economic growth because they are non-rival: once created, the same idea, design, or technological recipe can be used repeatedly and by many actors at once, unlike physical goods [**37**]. His work helped to move economic growth theory from a focus on capital accumulation alone toward a recognition that purposeful investment in knowledge, research and technological innovation can generate sustained growth.

The present theory supports those insights but shifts the focus of analysis. Romer's central concern was how ideas enter macroeconomic growth models and how institutions and incentives affect their production and diffusion. Instead of this, ideometrics asks how individuals, institutions and societies generate, evaluate, prioritise, implement and validate competing ideas under uncertainty. In this sense, ideometrics does not contradict Romer's theory, but rather offers to extend it in two directions: upstream, by examining how candidate ideas are generated and selected before they even contribute to growth; and downstream, by evaluating whether selected ideas truly produce realised future value. The novelty of ideometrics-based theory does not lie in suggesting that ideas matter for progress, which is already well established in endogenous growth theory, but rather in proposing a framework for measuring the quality of idea-selection processes themselves.

The Ideometric Index of Human Progress (IIHP) could become practically valuable contribution of this theory in the future, but the multiplicative form of its equation assumes that each component is measured on compatible scales and that the components interact multiplicatively rather than additively, or perhaps hierarchically. This assumption will need to be justified, or alternatively presented, through its first approximations based on future empirical analyses. The imminent next task for this theory is to compute a minimal viable real-world IIHP, using the existing examples from the stock market, start-up support and research grants funding. If this can be practically done, even imperfectly, then human progress becomes measurable as improvement in decision-making under uncertainty about ideas.

This paper also suggests that human progress itself may be understood as the progressive improvement of collective decision-making under uncertainty. Civilisations seem to advance when societies become increasingly accurate in identifying and implementing ideas of genuinely high future value. There is a potentially strong and interesting connection between ideometrics-based understanding of long historic cycles of civilisational success and failure, and the work by other two scientists: Samuel Scheffler focuses on the existential dependence of meaning on the existence of human civilisation in the future [**47**], William MacAskill's "longtermism" approaches the importance of future civilisation mainly through moral philosophy and existential risk reduction [**48**], while ideometrics aims to explain the informational dynamics that allow civilisation to survive and advance at all.

Therefore, we propose a possible new framework for understanding how societies allocate scarce capacity, time, energy, and resources toward competing ideas under uncertainty.

Although our transdisciplinary framework spans many different disciplines, it demonstrates the links between ideometrics and economics. Economics fundamentally studies allocation of scarce resources, decision-making under uncertainty, incentives, information, and societal welfare. This theory sees human progress as the increasing efficiency with which societies allocate scarce resources toward high-value ideas, which is an aim that is fundamentally economic in nature.

It is important to us to open the entire ideometrics-based theory of human progress to constructive criticism, revisions, and changes. We are opening a large space for discussion. Ideometrics, for now, is an emerging field of science based on a set of rational and internally consistent hypotheses, but they still require empirical confirmation. It actively invites comments, challenges, testing, ideas for further use, and alternative interpretations. The entire project could be viewed as an "open theoretical framework" that expects to evolve through peer criticism, open and constructive debate, and empirical work.

This should assist ideometrics to develop into a robust and widely used framework for measuring and improving the quality of idea selection under uncertainty, aiming to demonstrably improve economic and societal outcomes. Empirical demonstrations in domains such as stock market, start-up company investments, grant research funding, established company strategy, and public policy will be required, along with broad adoption by governments, investors, and institutions, followed by demonstrable improvement in real-world decision-making and welfare. Whether that occurs will not primarily depend on the "elegance" of this theory, but on the ability of the individuals and groups who recognise its potential value, adopt and implement it, to turn it into a rigorous, testable, and practically valuable scientific discipline.

## ACKNOWLEDGEMENTS

**Acknowledgement:** IR wishes to thank his wife Tonkica Rudan for continuing intellectual stimulation that assists him in conceptualising and drafting his creative work. IR and SK also thank Professor Stjepan Oreskovic from the University of Zagreb for the first feedback on the paper, for useful links to the work by William MacAskill and by Samuel Scheffler, and for suggesting that the work could be presented as "open theoretical framework" that invites constructive feedback.
**Funding:** This paper received no specific funding.
**Authorship contributions:** IR wrote the first completed draft of this paper. SK then provided further important intellectual content from the economics perspective, particularly on expected utility theory and decision theory, and assisted in producing more formal links with economics. Artificial Intelligence-based Large Language Model (ChatGPT 5.2) was used as an additional layer of checking for the novelty, accuracy and grammar of the key parts of the text.